\documentclass[a4,printer]{aa}
\usepackage{txfonts}
\usepackage{graphicx}

\begin{document}
   \title{(309239) 2007 RW$_{10}$: a large temporary quasi-satellite of Neptune} 
   \author{C. de la Fuente Marcos
           \and
           R. de la Fuente Marcos}
   \authorrunning{C. de la Fuente Marcos \and R. de la Fuente Marcos}
   \titlerunning{A quasi-satellite of Neptune}
   \offprints{C. de la Fuente Marcos, \email{nbplanet@fis.ucm.es}
              }
   \institute{Universidad Complutense de Madrid,
              Ciudad Universitaria, E-28040 Madrid, Spain}
   \date{Received 2 July 2012 / Accepted 7 September 2012}

   \abstract
      {Upon discovery, asteroid (309239) 2007 RW$_{10}$ was considered
       a Neptune Trojan candidate. The object is currently listed by the 
       Minor Planet Center as a Centaur but it is classified as a 
       Scattered Disk or Trans-Neptunian Object by others. Now that its 
       arc-length is 8,154 d and has been observed for more than 20 yr, a 
       more robust classification should be possible. 
       }  
      {Here we explore the orbital behaviour of this object in order to 
       reveal its current dynamical status.
       }
      {We perform $N$-body simulations in both directions of time to 
       investigate the evolution of its orbital elements. In particular, 
       we study the librational properties of the mean longitude.
       } 
      {Its mean longitude currently librates around the value of the 
       mean longitude of Neptune with an amplitude of nearly 50$^{\circ}$ 
       and a period of about 7.5 kyr. Our calculations show that it has 
       been in its present dynamical state for about 12.5 kyr and it will 
       stay there for another 12.5 kyr. Therefore, its current state is 
       relatively short-lived. Due to its chaotic behaviour, the object 
       may have remained in the 1:1 mean motion resonance with Neptune 
       for several 100 kyr at most, undergoing transitions between the 
       various resonant states.  
       }
      {(309239) 2007 RW$_{10}$ is currently a quasi-satellite, the first
       object of this dynamical class to be discovered around Neptune. 
       With a diameter of about 250 km, it is the largest known 
       co-orbital in the Solar System. Although it is not a Centaur now, 
       it may become one in the future as it appears to move in an 
       unstable region. Its significant eccentricity (0.30) and 
       inclination (36$^{\circ}$), strongly suggest that it did not form 
       in situ but was captured, likely from beyond Neptune. With an
       apparent magnitude of 21.1 at opposition (October), it is well 
       suited for spectroscopic observations that may provide information 
       on its composition and hence eventually its origin. 
       }


   \maketitle

   \section{Introduction}
      Minor planet (309239) 2007 RW$_{10}$ was discovered by the Palomar Distant Solar System Survey on September 9, 2007 (Schwamb 
      et al. 2007; Schwamb et al. 2010) and reobserved multiple times soon after (Schwamb et al. 2007; Parker et al. 2008). In 
      addition, a number of precovery images of the object were unveiled: it first appears in images obtained as part of the 
      Digitized Sky Survey (DSS) on June 1988 and 1990 from Palomar Mountain and it was pictured again in October 2001 and 
      September 2002 on behalf of the Near-Earth Asteroid Tracking (NEAT) project at Palomar. All this observational material 
      enabled the computation of a very reliable orbit characterized by both significant eccentricity (0.30) and inclination 
      (36$^{\circ}$). Herschel-PACS observations indicate that the albedo of the object is 8.3$^{+6.8}_{-3.9}$\% and its absolute 
      magnitude is $H_v$ = 6.39$\pm$0.61 which translates into a diameter of 247$\pm$30 km (Santos-Sanz et al. 2012). The 
      dynamical status of this object remains controversial. Upon discovery, it was considered a Neptune Trojan 
      candidate\footnote{http://www.boulder.swri.edu/ekonews/issues/past/n055/html/index.html} but it was reclassified as Centaur 
      shortly afterwards\footnote{http://www.boulder.swri.edu/ekonews/issues/past/n056/html/index.html}. It is currently listed 
      by the Minor Planet Center (MPC) as a Centaur but it has been classified as a Scattered Disk Object by Santos-Sanz et al. 
      (2012) and the JPL Solar System Dynamics portal includes this asteroid among the Trans-Neptunian Objects. Now that its 
      arc-length is 8,154 d and the object has been observed for more than 20 yr, a more robust dynamical classification should be 
      possible.

      In this Letter, we use $N$-body simulations to study the librational properties of the principal resonant angle of (309239) 
      2007 RW$_{10}$ with Neptune in order to understand its current dynamical status. The numerical model is described in Section 
      2; the results are presented in Section 3 and the long-term orbit behaviour is studied in Section 4. Our conclusions are 
      summarized in Section 5 after the corresponding discussion.

   \section{Numerical computations}
      For accurate initial positions and velocities we used the Heliocentric ecliptic Keplerian elements and their uncertainties 
      provided by the JPL\footnote{http://ssd.jpl.nasa.gov/sbdb.cgi} and the AstDyS-2 portal\footnote{http://hamilton.dm.unipi.it/astdys/index.php?pc=0} 
      (see Table \ref{elements}) and initial positions and velocities based on the DE405 planetary orbital ephemerides (Standish 
      1998)\footnote{http://ssd.jpl.nasa.gov/?planet\_pos} referred to the barycentre of the Solar System. The numerical 
      simulations were completed using a Hermite integration scheme (Makino 1991; Aarseth 2003); more details can be found in de 
      la Fuente Marcos \& de la Fuente Marcos (2012). Additional calculations were performed using the time-symmetric Hermite 
      method described by Kokubo et al. (1998) but it was found that, for the problem studied here, its overall performance was 
      lower and the results almost identical. The standard versions of these direct $N$-body codes are publicly available from the 
      IoA web site\footnote{http://www.ast.cam.ac.uk/$\sim$sverre/web/pages/nbody.htm}. These versions have been modified in order 
      to study the orbital evolution of (309239) 2007 RW$_{10}$. Our calculations include the perturbations by eight major planets 
      and treat the Earth and the Moon as two separate objects, they also include the three largest asteroids and the barycentre 
      of the dwarf planet Pluto-Charon system. Orbits are calculated forward and backward in time. In addition to the calculations 
      completed using the nominal orbital elements in Table \ref{elements} we have performed 100 control simulations using sets of 
      orbital elements sprinkled from the nominal ones within the accepted uncertainties (3$\sigma$) following a Monte Carlo approach. 
%
%------------------------------------------------------------------------------------------------------ Orbital elements 2007 RW10
%
         \begin{table}
          \fontsize{8}{11pt}\selectfont
          \tabcolsep 0.35truecm
          \caption{Heliocentric Keplerian orbital elements of (309239) 2007 RW$_{10}$ used in this research.
                   Values include the 1-$\sigma$ uncertainty.
                   (Epoch = JD2456200.5, 2012-Sep-30.0; J2000.0 ecliptic and equinox.
                   Source: JPL Small-Body Database and AstDyS-2.)
                  }
          \begin{tabular}{ccc}
           \hline
            semi-major axis, $a$                        & = & 30.323$\pm$0.005 AU \\
            eccentricity, $e$                           & = & 0.29957$\pm$0.00007 \\
            inclination, $i$                            & = & 36.06825$\pm$0.00014 $^{\circ}$ \\
            longitude of the ascending node, $\Omega$   & = & 187.03214$\pm$0.00007 $^{\circ}$ \\
            argument of perihelion, $\omega$            & = & 96.734$\pm$0.013 $^{\circ}$ \\
            mean anomaly, $M$                           & = & 58.95$\pm$0.02  $^{\circ}$ \\ 
           \hline
          \end{tabular}
          \label{elements}
         \end{table}
%
%---------------------------------------------------------------------------------------------------------------------------------
%
%
%---------------------------------------------------------------------------------------------------------------------------------
%
     \begin{figure}
       \centering
        \includegraphics[width=\linewidth, height=5.0cm]{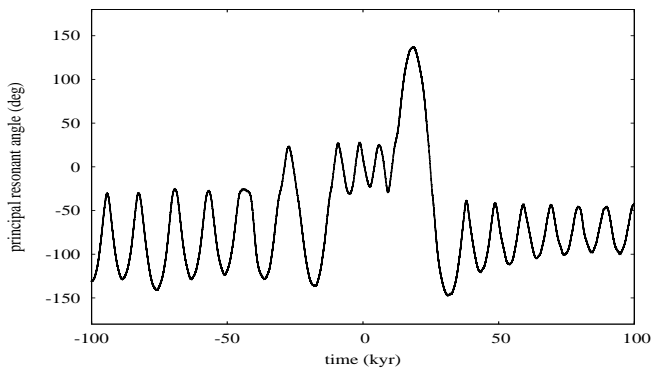}
        \caption{Evolution of the main resonant angle, $\lambda_{r}$ = $\lambda$ - $\lambda_N$, of (309239) 2007 RW$_{10}$ 
                 over time. It librates around $0^{\circ}$ in the time interval $\sim$(-12.5, 12.5) kyr. In all figures, the 
                 zero of time is JD2456200.5.
                }
        \label{sigma}
     \end{figure}
%
%---------------------------------------------------------------------------------------------------------------------------------
%

   \section{(309239) 2007 RW$_{10}$: current status}
      Our main objective is the study of the librational properties of the principal resonant angle of (309239) 2007 RW$_{10}$ in 
      order to reasses its current dynamical class. The principal resonant angle of the object is given by $\lambda_{r}$ = 
      $\lambda$ - $\lambda_N$, where $\lambda$ is the mean longitude of the asteroid and $\lambda_N$ is the mean longitude of 
      Neptune ($\lambda$ = $M$ + $\Omega$ + $\omega$, $M$ is the mean anomaly, $\Omega$ is the longitude of the ascending node and 
      $\omega$ is the argument of perihelion). If it librates around 0$^{\circ}$ we have the quasi-satellite state, this is a 
      specific configuration of a 1:1 mean motion resonance, one in which the body librates around the mean longitude of its 
      associated planet; the minor planet orbits the Sun in an approximate ellipse with the same (mean) period as the planet. When 
      viewed in a frame of reference that corotates with the planet, the quasi-satellite follows a retrograde path around the body 
      over the course of an orbital period. Such motion is stabilized by the host planet. The stability of quasi-satellite orbits 
      has been studied by Mikkola et al. (2006). Although brought to the attention of the astronomical community by Mikkola \& 
      Innanen (1997), the concept behind the term quasi-satellite was first studied by Jackson (1913) and the associated 
      energetics was first discussed by H\'enon (1969). If the principal resonant angle librates around the values +60$^{\circ}$ 
      or -60$^{\circ}$ (or 300$^{\circ}$), the object is called a Trojan and the associated path, a tadpole orbit. If the 
      libration amplitude is larger than 180$^{\circ}$, the path is called a horseshoe orbit. Recurrent transitions between 
      resonant states are possible (quasi-satellite, tadpole, horseshoe) for objects with both large eccentricity and inclination 
      (Namouni et al. 1999). For example, a compound orbit between the Trojan and quasi-satellite states is also called a 
      large-amplitude Trojan when the libration amplitude is less than 180$^{\circ}$.  
%
%---------------------------------------------------------------------------------------------------------------------------------
%
     \begin{figure}
       \centering
        \includegraphics[width=\linewidth]{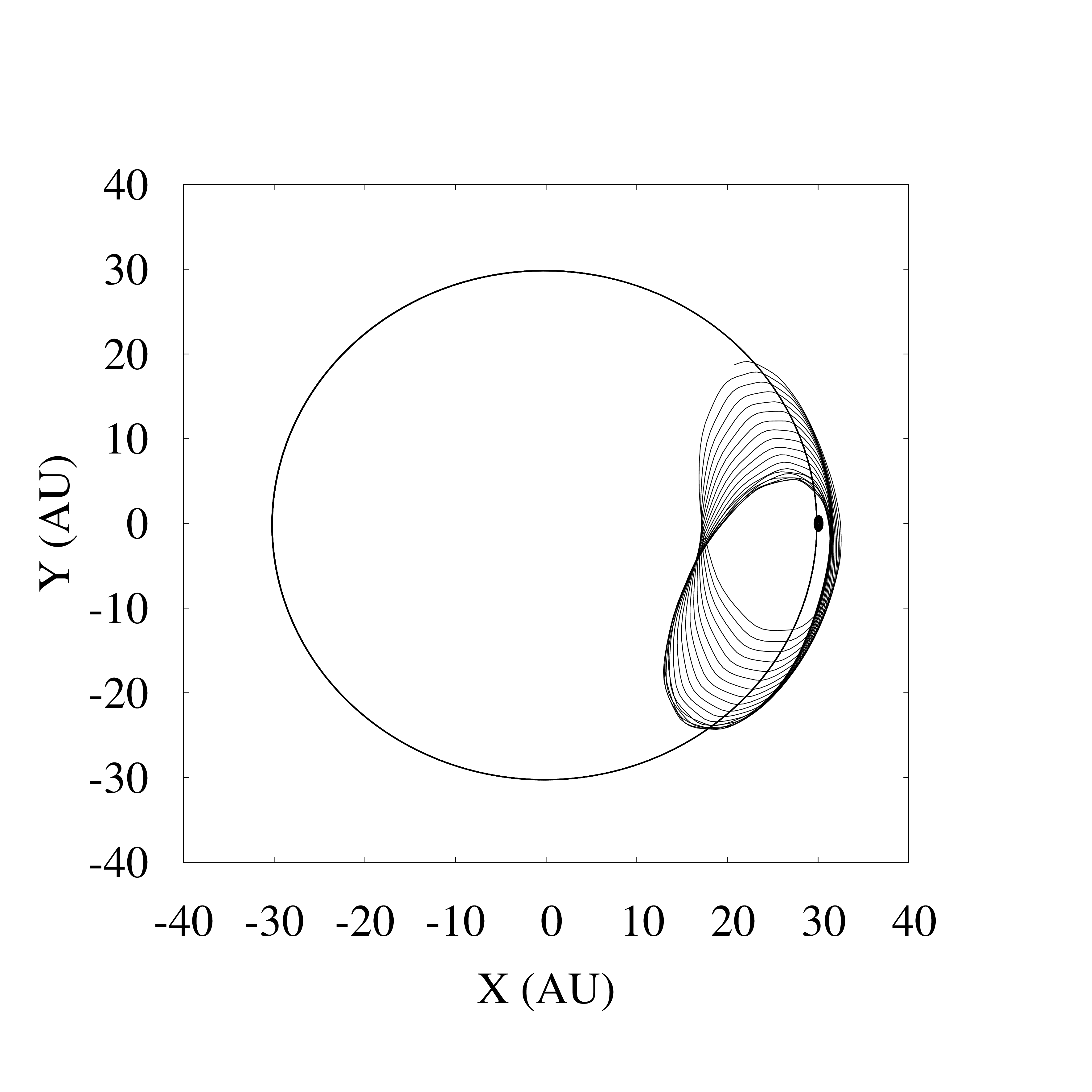}
        \caption{The motion of (309239) 2007 RW$_{10}$ over the next 3,000 yr in a coordinate 
                 system rotating with Neptune. The quasi-satellite appears to follow a 
                 precessing kidney-shaped path when viewed from Neptune. The orbit and the 
                 position of Neptune are also indicated. 
                }
        \label{qs}
     \end{figure}
%
%---------------------------------------------------------------------------------------------------------------------------------
%

      In Fig. \ref{sigma} we plot $\lambda_{r}$ for the nominal orbit in Table \ref{elements}. The principal resonant angle 
      currently librates around 0$^{\circ}$. In Celestial Mechanics, this is the characteristic property of an object in the 
      quasi-satellite dynamical state. In Fig. \ref{qs} we plot the motion of the object for the next 3,000 yr in a coordinate 
      system that rotates with Neptune. In the rotating coordinate system, the quasi-satellite follows a precessing kidney-shaped 
      retrograde path when viewed from Neptune. Our control calculations produce similar results for the time interval (-13, 24) 
      kyr (see Fig. \ref{all}B). Therefore, (309239) 2007 RW$_{10}$ is a quasi-satellite of Neptune; it has remained as such for 
      about 12.5 kyr and it will remain in that state for another 12.5 kyr librating around 0$^{\circ}$ with amplitude 
      40$^{\circ}$-60$^{\circ}$ and a period of about 7.5 kyr. Here by amplitude we mean the difference between the maximum and 
      the minimum values of $\lambda_{r}$ in a period. These values are similar to those quoted for Neptune Trojans (Zhou et al. 
      2009, 2011). (309239) 2007 RW$_{10}$ is a temporary quasi-satellite of Neptune which only survives 3 librations of the 
      resonant angle before leaving the state. The current trajectory followed by this object is fairly chaotic, its e-folding 
      time during the quasi-satellite phase is nearly 1 kyr; simulations can only reliably compute the motion of such an object 
      for a time interval not longer than about 50,000 yr. The object is clearly non-primordial. The presence of quasi-satellites 
      around Neptune was predicted by Wiegert et al. (2000); in their paper, it was concluded that the outermost planets (Uranus 
      and Neptune) could be the preferred locations to find objects in the quasi-satellite dynamical state.  

   \section{Long-term orbit behaviour: quasi-satellite or Trojan}
\begin{figure}
       \centering
        \includegraphics[width=\linewidth, height=15.5cm]{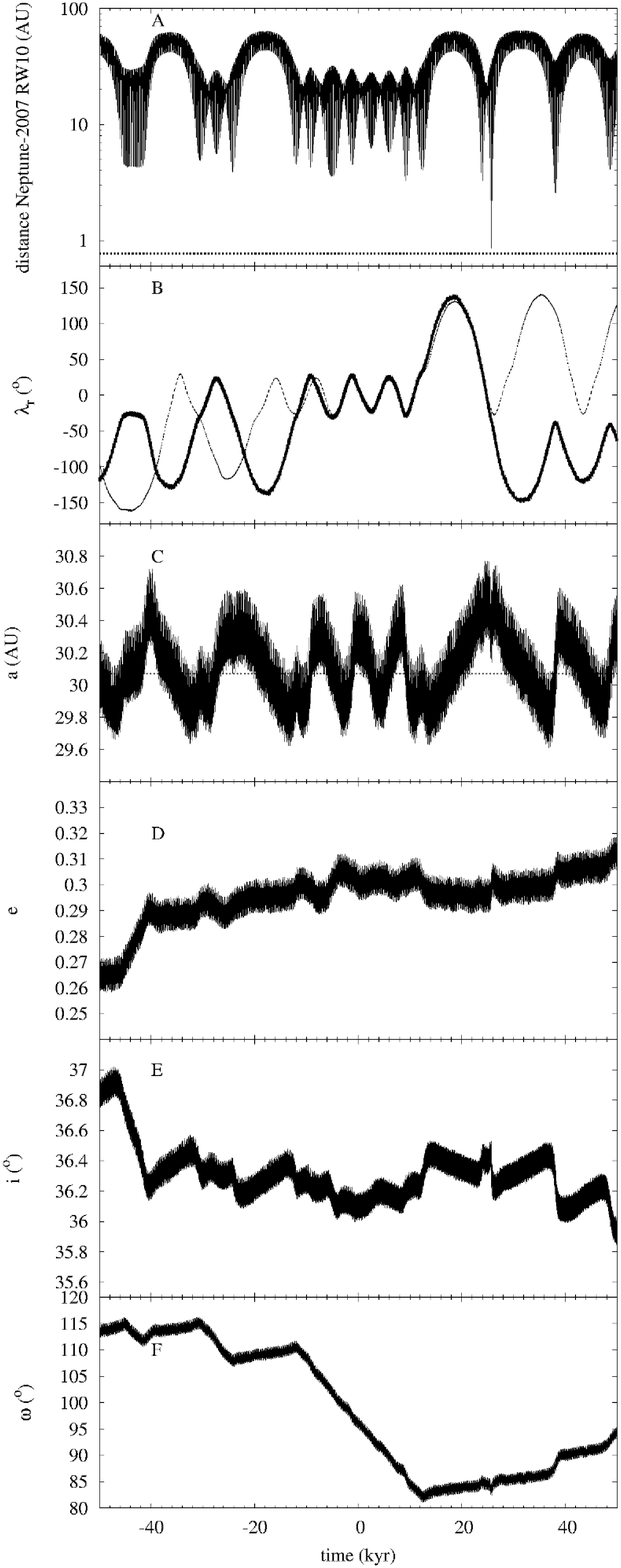}
        \caption{Time evolution of various parameters. The distance of (309239) 2007 RW$_{10}$ from Neptune (panel A); the
                 value of the Hill sphere radius of Neptune, 0.769 AU, is displayed. The resonant angle, $\lambda_{r}$ (panel B) 
                 for the nominal orbit in Table \ref{elements} (thick line) and one of the control orbits (thin line). This 
                 particular control orbit has been chosen close to the 3-$\sigma$ limit so its orbital elements are most different 
                 from the nominal ones. The orbital elements $a$ (panel C) with the current value of Neptune's semi-major axis, 
                 $e$ (panel D) and $i$ (panel E). The argument of perihelion (panel F) evolves as predicted by Namouni (1999).  
                }
        \label{all}
     \end{figure}

\begin{figure}
       \centering
        \includegraphics[width=\linewidth, height=5.0cm]{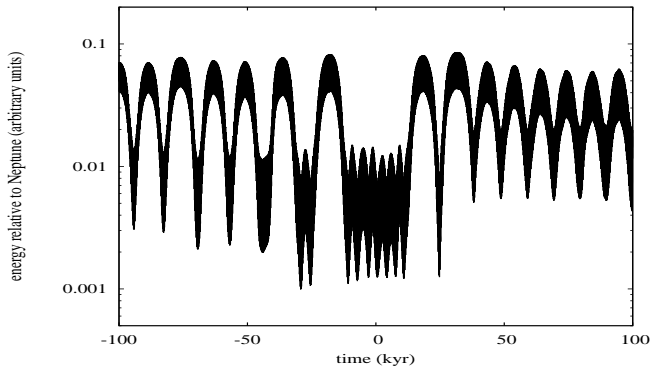}
        \caption{Total energy (specific orbital energy) of (309239) 2007 RW$_{10}$ relative to Neptune. The quasi-satellite state,
                 even if not bound (energy $<$ 0), is clearly less energetic than the other resonant states.
                }
        \label{energy}
     \end{figure}

      Our calculations suggest that the present state of the orbit has lasted for about 12.5 kyr but half the control calculations 
      increase that number to 20 kyr (see Fig. \ref{all}B). About 12.5 kyr from now, the asteroid will become an L$_4$ 
      large-amplitude Trojan with $\lambda_{r}$ librating around +60$^{\circ}$. In the past, the object used to be a Trojan around 
      the L$_5$ Lagrangian point. The distance of (309239) 2007 RW$_{10}$ from Uranus remains larger than 4 AU during the 
      displayed time interval suggesting that encounters with Uranus do not cause the asteroid to depart from the quasi-satellite 
      orbit in either direction of time. However, the distance to the object from Neptune can become as small as 0.86 AU (see 
      Figure \ref{all}A), close to the Hill distance for Neptune that is 0.77 AU. Much closer approaches (even under 0.1 AU) have 
      been observed both in the main simulation and in the control calculations but they fall outside the time range in the 
      figures. Figure \ref{all} (all panels) strongly suggests that the transitional resonant behaviour observed is driven by 
      Neptune as the timings of the various transitions coincide with close encounters between the asteroid and the giant planet. 
      In Fig. \ref{all}C, D, E we illustrate the temporal evolution of the nominal orbit. However, the key parameter here is the 
      argument of perihelion (see Fig. \ref{all}F); during the quasi-satellite episodes (even if brief) its value decreases 
      uniformly (in our case, at a rate of $\dot{\omega} = -0.001^{\circ}$/yr) as predicted by Namouni (1999) but when the object 
      follows a tadpole or horseshoe path, its value increases. In the Solar System and for a minor body moving in an inclined 
      orbit, close encounters with major planets are only possible in the vicinity of the nodes. Transfers between 
      quasi-satellite, horseshoe and tadpole orbits are the result of the libration of the nodes (Wiegert et al. 1998). On the 
      other hand, the quasi-satellite dynamical state is characterized by a value of the relative specific orbital energy (see 
      Fig. \ref{energy}) that is close to the binding border.  
%
%---------------------------------------------------------------------------------------------------------------------------------
%
%
%---------------------------------------------------------------------------------------------------------------------------------
%

%
%---------------------------------------------------------------------------------------------------------------------------------
%
%
%---------------------------------------------------------------------------------------------------------------------------------
%
   \section{Discussion and conclusions}
      From a dynamical standpoint, our calculations indicate that (309239) 2007 RW$_{10}$ is currently a quasi-satellite of 
      Neptune. The object has remained in its current dynamical state for about 12.5 kyr and will continue following that type of 
      orbit for about 12.5 kyr into the future. Although temporary, its present status as quasi-satellite is very robust given the 
      current level of accuracy of the ephemeris and how consistent our results in the time interval (-13, 24) kyr are. On the 
      other hand, the object may have remained in the 1:1 mean motion resonance with Neptune for hundreds of kyr at most, 
      transitioning between the various co-orbital resonant states, as it inhabits a very chaotic region and moves in a 
      dynamically hot orbit. Its significant eccentricity (0.300) is among the largest for any co-orbitals (only Jupiter Trojans 
      like 2010 FH$_{60}$, 0.300, or 2012 BZ$_{7}$, 0.299, appear to have comparable values) and its high inclination 
      (36$^{\circ}$) is the largest among Neptune co-orbitals. These properties and our own results do not favour an scenario in 
      which this object is part of a primordial population of Neptune co-orbitals or the result of collisional processes in the 
      vicinity of Neptune. The object is probably an extreme dynamical relative of the L$_5$ Neptune Trojan 2004 KV$_{18}$, which 
      is believed to have been originated beyond the orbit of Neptune (Horner \& Lykawka 2012). The origin of temporary 
      co-orbitals has recently been discussed by, e.g., Lykawka et al. (2011), Petit et al. (2011) and Gladman et al. (2012) and 
      references inside.

      Quasi-satellites have been found around Venus (Mikkola et al. 2004), Earth (Wiegert et al. 1997; Connors et al. 2002; 
      Connors et al. 2004; Brasser et al. 2004; Christou \& Asher 2011), the dwarf planet (1) Ceres and the large asteroid (4) 
      Vesta (Christou 2000; Christou \& Wiegert 2012), Jupiter (Kinoshita \& Nakai 2007; Wajer \& Kr\'olikowska 2012) and Saturn 
      (Gallardo 2006). Quasi-satellite orbits around Uranus and Neptune have been predicted to be stable for up to 1 billion years 
      (Wiegert et al. 2000) but none have been identified in that region until now. (309239) 2007 RW$_{10}$ is the first {\it bona 
      fide} quasi-satellite found around Neptune; as a co-orbital, it adds to the 8 Trojans previously discovered. It is also the 
      largest known object in the 1:1 mean motion resonance with any major planet. The previous record holder was the largest of 
      Jupiter's Trojans, 624 Hektor with a diameter of 203.0$\pm$3.6 km (Fern\'andez et al. 2003). With an apparent visual 
      magnitude of 21.1 when at opposition in October and in contrast with known Neptune Trojans, this quasi-satellite is bright 
      enough to be studied spectroscopically to investigate its surface composition. This will provide definite clues on its 
      origin and past evolution.

   \begin{acknowledgements}
      The authors would like to thank S. Aarseth for providing the codes used in this research and two anonymous referees for 
      a number of helpful suggestions which improved the overall presentation of this Letter. This work was partially supported by 
      the Spanish 'Comunidad de Madrid' under grant CAM S2009/ESP-1496 (Din\'amica Estelar y Sistemas Planetarios). We thank Dr. 
      Mar\'{\i}a Jos\'e Fern\'andez-Figueroa, Dr. Manuel Rego Fern\'andez and the Department of Astrophysics of Universidad 
      Complutense de Madrid (UCM) for providing excellent computing facilities. Most of the calculations and part of the data 
      analysis were completed on the 'Servidor Central de C\'alculo' of the UCM and we thank Santiago Cano Als\'ua for his help 
      during that stage. In preparation of this Letter, we made use of the NASA Astrophysics Data System and the ASTRO-PH e-print 
      server.  
   \end{acknowledgements}

\end{document}